\documentclass{egpubl}
\usepackage{egsr2026}
\usepackage{todonotes}
\usepackage{multirow}
\usepackage{colortbl}
\usepackage{amsmath}
\usepackage{amssymb}
\usepackage{enumitem}
\definecolor{RoyalPurple}{RGB}{128, 0, 128}
\definecolor{RoyalBlue}{RGB}{65, 105, 225}
\definecolor{EmeraldGreen}{RGB}{0, 128, 0}
\definecolor{AmberOrange}{RGB}{255, 191, 0}

\providecommand{\Description}[1]{}

\newcommand{\wi}{\mathbf{\omega_i}}
\newcommand{\wo}{\mathbf{\omega_o}}

\newcommand{\R}{\mathbb{R}}

\newcommand{\fc}{f_{\text{c}}}
\newcommand{\fg}{f_{\text{g}}}
\newcommand{\ft}{f_{\text{t}}} %
\newcommand{\fa}{f_{\text{a}}}

\newcommand{\nn}{\psi} %
\newcommand{\p}{w} %
\newcommand{\kd}{k_d} %

\newcommand{\latent}[1]{$\scriptscriptstyle{l=#1}$}

\newcommand{\SMAPE}{SMAPE} %

\CGFStandardLicense

\usepackage[T1]{fontenc}
\usepackage{dfadobe}  
\usepackage{cite}  %
\BibtexOrBiblatex
\electronicVersion
\PrintedOrElectronic

\usepackage{egweblnk} 

\newcommand{\prepeerreviewmark}{%
  \makebox[0pt][l]{%
    \raisebox{112pt}[0pt][0pt]{%
      \parbox[t]{\textwidth}{%
        \raggedleft\normalfont\sffamily\fontsize{8.2}{9.2}\selectfont\color{black!65}%
       Preprint version, prior to peer review. A revised version of this paper has been accepted to the Eurographics Symposium on Rendering (EGSR) 2026. \href{https://diglib.eg.org/handle/10.1111/cgf70540}{Link}.

      }%
    }%
  }%
}

\title[A Hybrid Neural-Microfacet BRDF Model for Real-Time Rendering]%
      {A Hybrid Neural-Microfacet BRDF Model for Real-Time Rendering}

\author[L. De Oliveira et al.]{
  L. De Oliveira$^{1,2}$,
  A. Karpova$^{1}$,
  G. Nader$^{1}$,
  A. Houdard$^{1}$,
  P. Mézières$^{2}$,
  D. Rioux-Lavoie$^{1}$,
  R. Pacanowski$^{2}$
  \\
  $^{1}$Ubisoft La Forge,
  $^{2}$Inria
}

\begin{document}

\teaser{
 \includegraphics[width=1\linewidth]{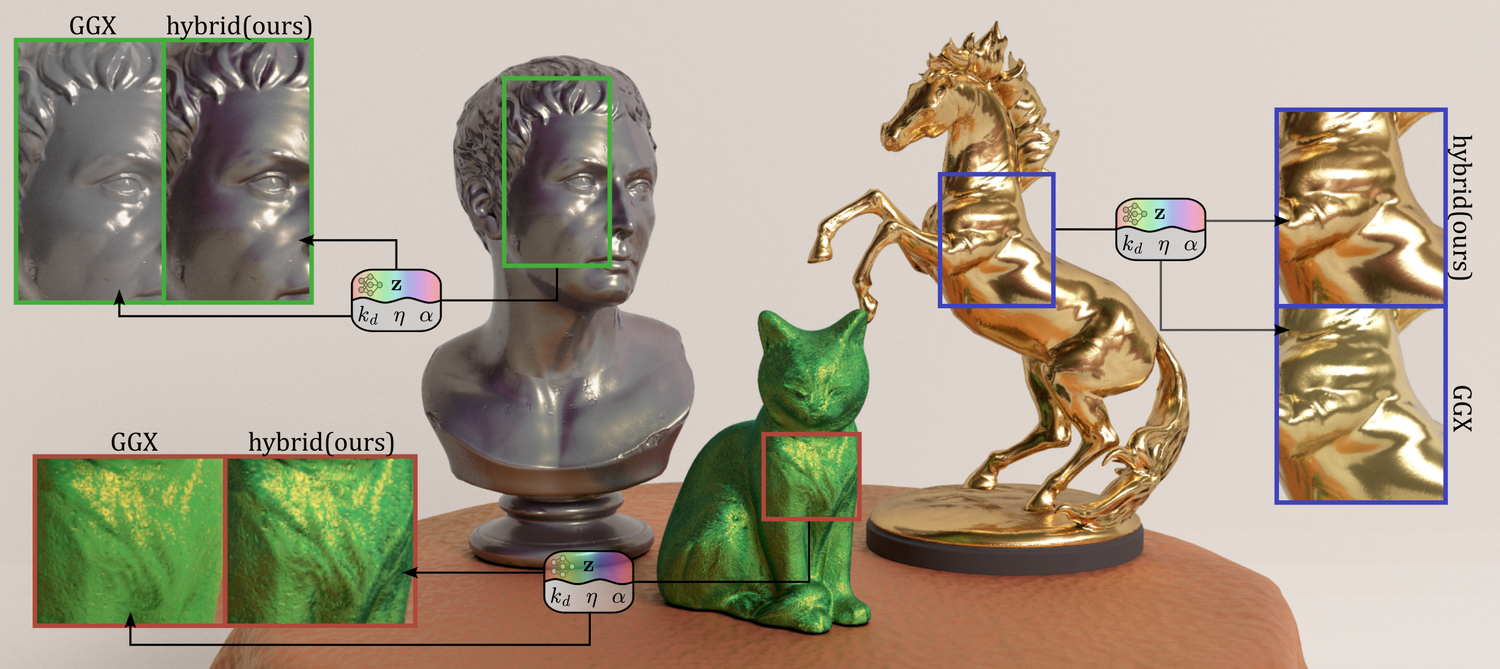}
 \centering
  \caption{
  Our hybrid model is fitted from measured BRDFs and represented as a compact set of microfacet parameters $(\kd, \eta, \alpha)$ and a low-dimensional latent code $\mathbf{z}$ shared across a single neural network. 
  For each material, the neural correction improves upon the microfacet model alone, capturing subtle appearance effects, while the analytical component remains a faithful and useful approximation. 
  From left to right \emph{irid\_flake\_paint2} on the bust, \emph{cc\_green\_malachite} on the cat and \emph{aniso\_metallic\_paper\_gold} on the horse from the RGL database~\cite{dupuAdapt18}.
  }
}
\label{fig:teaser}

\maketitle


\begin{abstract}
Over the past decade, microfacet-based 
BRDF models have formed the foundation of real-time rendering pipelines. 
Despite their widespread use, they often fail to reproduce subtle appearance effects arising from complex light–surface interactions, which have led to the emergence of specialized physics-based models for specific optical phenomena (e.g., diffraction, iridescence, multilayers). Although more accurate, these models lose versatility and lack performance for real-time rendering. 
Recently introduced, neural models have demonstrated their ability to approximate BRDF 
reference data coming from measurements, simulations, or even complex shading networks. 
However, most current neural models require relatively large networks, making them
costly for real-time rendering.
In this paper, we introduce a hybrid model that combines a GGX-type microfacet model and a neural model to leverage the best features of both representations.
The neural component corrects the appearance approximated by the microfacet component, allowing much smaller network than in existing neural models. %
We show that, at identical memory cost, our model approximates measurements better than state-of-the-art neural models for a low evaluation overhead compared to a microfacet-based model. %
Furthermore, our hybrid model remains easily editable by artists and benefits from an important sampling scheme, making it attractive for both offline and real-time rendering.


\end{abstract}

\section{Motivation}

The appearance of an object emerges from the interaction between the light and its surface. 
Modeling it accurately is at the core of computer graphics and remains an active research topic. The reflection of the light by the surface is %
described by a physical quantity known as the Bidirectional Reflectance Distribution Function (BRDF). %
In computer graphics, especially in a real-time context, an ideal BRDF representation should satisfy the following properties:

\begin{itemize}[align=parleft] %
\item[(i)] \hspace{0.3cm}\emph{expressiveness}, to represent a large number of materials, 
\item[(ii)] \hspace{0.3cm}\emph{robustness} when fitting measurements or simulated data,
\item[(iii)] \hspace{0.3cm}\emph{fast evaluation} with importance sampling support, 
\item[(iv)] \hspace{0.3cm}\emph{low memory impact} for stored parameters,
\item[(v)] \hspace{0.3cm}\emph{editability} through meaningful parameters.  %
\end{itemize}

Over the past decade, microfacet-based  BRDF models have formed the foundation of real-time rendering pipelines ~\cite{RTR4}. 
Despite their widespread use, they often fail to reproduce subtle appearance effects arising from complex light–surface interactions. This has motivated specialized physics-based models for specific optical phenomena (e.g., diffraction, iridescence, multilayers). 
Although more accurate, these models lose versatility and generally lack performance for real-time rendering.  
Recently, neural models have shown strong ability to approximate BRDF reference data coming from measurements, simulations, or complex shading networks. 
However, most current neural models require relatively large networks, making them costly for real-time rendering. 
Furthermore, parameters of a neural model are not meaningful in terms of appearance control, which makes them less suitable for authoring compared to analytical BRDF formulations.

In this paper, we introduce a hybrid model that combines an analytical microfacet-based model %
and a neural model to leverage the best features of each part. We then demonstrate that such a model has good characteristics regarding the above-mentioned requirements. %
The neural component acts as a corrective term, refining the output of the analytic BRDF to capture appearance effects that are otherwise difficult to reproduce.
To keep memory tight \textbf{(iv)}, we enforce the learning of a common neural network for a whole collection of measured BRDFs, and only differentiate the correction term through a low-dimensional latent space as input of the network. 
Furthermore, we  propose a joint training strategy that learns the parameters of both the analytic and neural components simultaneously on the whole BRDF collection \textbf{(ii)}, and demonstrate high-quality reconstruction \textbf{(i)} of many measured BRDFs from three databases (MERL~\cite{matuDatad03}, UTIA~\cite{filip14template}, RGL~\cite{Dupuy2018Adaptive}).
We also enforce the reconstruction of the microfacet component to be close to the reference data during the training, implying that the network only learns residual effects rather than duplicating the base model. 
Restricting the network to predict residuals allows us to use a shallow multilayer perceptron (MLP) that remains small and computationally efficient \textbf{(iii)}, enabling high visual fidelity with minimal computational overhead, while still allowing the parameters of the analytic BRDF to be changed \textbf{(v)}.
The reconstruction accuracy achieved by our model matches existing neural approaches while relying on a significantly smaller neural network, and our model can even outperform them when operating at comparable network sizes. 
As a result, it delivers clear improvements over standard analytical BRDF models while preserving real-time performance.

In the following, we review the relevant previous work on BRDF representations (Section~\ref{sec:prev}), then introduce our hybrid model and the joint fitting procedure (Section~\ref{sec:hybridBRDF}). We then demonstrate its expressiveness capabilities by approximating a large number of real-world measurements (Section~\ref{sec:results}).
More precisely, we evaluate our model in terms of BRDF reconstruction error and rendering error, comparing it to the neural state-of-the-art approach~\cite{zeltRealt24} (Sections~\ref{sec:quality} and \ref{sec:renderingPerf}), and we perform ablation studies to validate our formulation (Section~\ref{sec:ablation}). Finally, we discuss the advantages of our hybrid formulation both for real-time rendering performance and intuitive authoring in Section~\ref{sec:discussion} and conclude by discussing  limitations and future directions (Section~\ref{sec:limitations}).

\section{Related Work \label{sec:prev}}

In the context of real-time rendering, building a representation that simultaneously satisfies the five criteria defined in the previous section remains particularly challenging. In the following, we review the main families of BRDF representations: tabulated approaches, analytic representations, non-parametric models, and neural ones.

\emph{Tabulated BRDFs} are obtained thanks to dense measurements (e.g.,~\cite{Matusik03}) or long light transport simulation (e.g.,~\cite{guoPosit18, gambEffic20,waveBRDFSimu23}) and permit very high quality reconstruction and fast evaluation. However, they are generally impractical in production rendering due to prohibitive memory requirements. A more practical approach regarding memory cost is to model the BRDF with a parametric function.

\emph{Single-Lobe BRDF Models}
are most commonly formulated using the microfacet framework, introduced to the Computer Graphics community by Cook and Torrance~\shortcite{cookRefle82}. 
This framework considers that the reflectance of a surface can be modeled as a collection of small Fresnel mirrors (i.e., microfacets) whose orientations  are statistically defined by a distribution term (cf.~\cite{waltMicro07}). 
Most of the time, the microfacet representation is supplemented by a Lambertian term. Although very efficient in terms of evaluation for real-time rendering~\cite{lagarde2014moving} and easy to manipulate through its 7 or 8  
parameters, the representation remains inherently limited in terms of expressiveness (cf.~\cite{Ngan2005}). 
To overcome these limitations, different types of extensions have been proposed, such as correction for multiple-scattering  between the microfacets (cf.~\cite{Heitz2016,kulla2017revisiting,FengHanrahan2018} or to encapsulating wave phenomena (e.g.,\cite{belcour2017practical,fourneau:hal-05126010,twoScaleHP17}). 
Although these approaches significantly enrich the appearance of materials, they are usually effect-specific and increase both the number of parameters and the overall complexity of handling different materials.
Another approach to augment expressiveness is to consider multiple-lobe parametric models.

\emph{Multiple-lobe BRDF Models} are either obtained empirically (e.g., \cite{burlPhysi12,Lafortune97}) or by modeling the surface as a stack of parallel interfaces or media layers to obtain so-called layered BRDFs.
The Weidlich and Wilkie~\shortcite{weidArbit07} model, optimized later by Elek~\shortcite{elek2010layered}, was one of the first  to provide an evaluation of the BRDF resulting from multiple layers.
These models, which do not take fully into account scattering effects, are improved by the recent introduction of statistical models \cite{dedineRende22,belcour2018efficient,yamaguchi2019real,weier2020rendering,randrianandrasana2021transfer}.
Despite these advances, the use of layered BRDF models remains limited since they require high memory demands (each layer requires its own set of parameters). 
In practice, to balance quality, performance, and storage costs, real-time systems typically restrict the number and type of layers that can be represented. 
Moreover, although these models significantly increase the range of achievable appearances,
the large number of interdependent parameters and the indirect relationship between model parameters and perceived appearance make intuitive control difficult.
In particular, layered BRDFs require an explicit material decomposition that is hard to define and even harder to adjust a posteriori, as modifications to one layer propagate non-linearly through the entire model.
Furthermore, as shown by Ngan~\shortcite{Ngan2005}, multiple-lobe models tend to be numerically unstable during the fitting process. For this reason, the aforementioned layered BRDF models are rarely tested against real-world BRDF measurements. Aside from parametric representations, non-parametric models have also been studied to represent BRDFs.

\emph{Non-parametric Models} 
overcome physical constraints and seek and consider the BRDF as a signal defined on the hemisphere that can be projected into basis functions (e.g., Spherical harmonics~\cite{kautz2002fast,soler2015efficient}, wavelets~\cite{ShroderWim1995,claustres2006}, Fourier basis~\cite{jakob2014comprehensive,zeltner2018layer}) or reconstructed after a PCA on a database (cf.~\cite{Matusik03}). Other approximating approaches use radial basis functions~\cite{svbrdfRBF05}, Spherical Gaussians~\cite{Wang2009,AnisoSG2013}, Zernike polynomials~\cite{zernikeBRDF96}, rational  functions~\cite{rationalFunctions2012}, Gaussian Mixtures~\cite{cooper2021estimating}, tensor product~\cite{TensorBRDF2011} or 
dictionaries~\cite{Tanaboon2022}.
These methods can represent all types of materials very precisely, as long as dense measurements or simulations are provided as input.
Their main limitations are their memory cost, because it grows quadratically with the specularity~\cite{Mahajan2008} of the materials, and their lack of editing capabilities.

\emph{Neural methods}
have emerged as a powerful alternative to analytic BRDF models, predicting reflectance directly using neural networks.
Such methods are effective in modeling even complex and layered materials \cite{zeltRealt24, fanNeura22} and can also work as compression for measured materials \cite{huDeepB20, sztrNeura21}.
The use of autoencoders and hypernetworks \cite{gokbHyper24} allows for exploration of learned manifolds of BRDFs, enabling interpolation between data points or edition of the BRDF model through the latent variables \cite{zhenCompa22}. 
Neural BRDF models do have some drawbacks compared to more physically based approaches: like most data-driven methods, they lack inductive bias for this task, thus requiring a large amount of data to train.
Furthermore, when dealing with deep neural networks, the inference cost can quickly become prohibitive for real-time rendering. 
However, it has been shown that the use of shallow MLPs can be evaluated in real-time during the rendering stage \cite{weinreich2024real,zeltRealt24,laurHardw25}. 
Recent methods, such as \cite{zeltRealt24, douRealT24}, explore the use of embeddings for real-time inference, reducing the size of the neural network but increasing the memory cost of the method. 
Other neural BRDFs methods tackle the issue of inference speed more directly. 
Xu et al.~\shortcite{xuComp25} apply Int8 quantization-aware training to achieve an order-of-magnitude speedup, while Xu et al.~\shortcite{xuMobile26} combine a coarse-to-fine network architecture with texture-space shading and spatiotemporal amortization to reach real-time frame rates on mobile VR devices.
Regardless of the inference strategy, efficient importance sampling for neural BRDFs remains an open problem~\shortcite{xuComp25}.
Existing solutions either require a dedicated network evaluated at every ray bounce~\cite{Bai2023,Wu2025}, or rely on a an expensive conversion to an analytical proxy \cite{zeltRealt24}

Across these families of BRDF representations, none simultaneously satisfies the ideal BRDF representation for real-time applications.
Our work aims to combine the best of both the analytical and neural worlds: by making the analytical component an integral part of the model, we retain the physical structure and importance sampling scheme of microfacet models while leveraging the expressiveness of neural networks to correct their limitations.

\begin{figure*}[ht]
  \includegraphics[width=\textwidth]{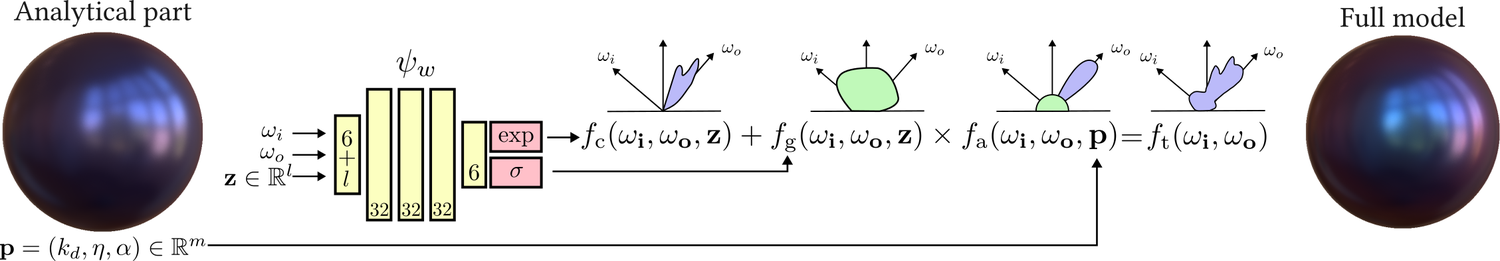}
  \caption{Overview of our method. We jointly optimize a latent code $\mathbf{z}$ and analytical parameters $\mathbf{p} = (\kd, \eta, \alpha)$ for each BRDF and a shared neural network. The output of the neural network (here, a $32\,\times 3\,$ MLP) is split into two parts: the correction $\fc(\wi,\wo,\mathbf{z})  \in \R_+^3$ with an exponential activation and the gate $\fg (\wi,\wo,\mathbf{z})\in [0,1]^3$ with a sigmoid activation. These outputs are combined with the evaluation of the analytical model to produce our improved BRDF.}
  \label{fig:method}
\end{figure*}

\section{Hybrid BRDF Model \label{sec:hybridBRDF}}

In this section, we introduce a hybrid BRDF model that combines a standard analytical reflectance formulation with learned neural components, leveraging the strengths of both physically based and neural approaches.
The idea is to let the analytical component model the dominant reflectance lobes and capture the primary perceptual characteristics of the material, while more complex and subtle effects are delegated to a neural network.
We jointly optimize both the parameters of the analytical BRDF and the neural network weights in an end-to-end training process, allowing the model to automatically distribute representational capacity between the analytical and learned components.

\subsection{General Formulation}
Our hybrid formulation models the target BRDF $\ft$ as the combination of an analytical component $\fa$ modulated by a gating function $\fg$ and corrected by an additive residual term $\fc$:
\begin{equation}\label{eq:formulation}
    \ft(\wi,\wo) = \fc(\wi, \wo, \mathbf{z}) + \fg(\wi, \wo, \mathbf{z}) \times \fa(\wi,\wo, \mathbf{p}),
\end{equation}
where $\wi$ and $\wo$ are respectively the incident light and viewing directions, $z\in\R^l$ denotes a per-BRDF latent code and $p\in\R^m$ the per-BRDF parameters of the analytical model.
The neural component of the model consists of both the gating function $\fg$ and residual term $\fc$, which are jointly predicted by a single neural network~$\nn_\p$:
\begin{equation}
    \left( \fc, \fg \right) = \nn_\p(\wi,\wo;z)\,,
\end{equation}
where $\p$ denotes the trainable parameters of the neural network. Figure~\ref{fig:method} provides an overview of our approach and illustrates the structure of the proposed hybrid BRDF representation.
We additionally enforce $\fc$ to be positive, and constrain $\fg \in [0, 1]$ using an exponential activation for the residual term and a sigmoid activation for the gating term. 
These constraints ensure that the final BRDF remains positive even for combinations of latent codes and analytical parameters not seen during training. This structural property enables intuitive material editing as further detailed in Section~\ref{sec:editability}.

In this hybrid formulation, each term plays a distinct role. The analytical term $\fa$ models the primary reflectance behavior and ensures that the model remains grounded to physically-based parameters. The gating function $\fg$ provides a direction-dependent modulation that adjusts the influence of the analytical component. Finally, the additive correction term $\fc$ enables the model to reproduce complex reflectance effects that the analytic model cannot express.

\subsection{Model Architecture \label{sec:archi}}

In practice, our hybrid model formulation requires defining both a neural network architecture $\nn$ and an analytical model $\fa$. In this work, we restrict ourselves to a single-lobe analytical model together with a lightweight MLP for the network to meet real-time rendering requirements.

\paragraph*{Analytical Model.}
Inspired by widely used models in real-time scenarios, we use an analytical component $\fa$ composed of a Lambertian lobe and a GGX~\cite{waltMicro07} microfacet-based specular lobe:
\begin{equation}
   \fa(\wi, \wo ; p) = \frac{\kd}{\pi} + \frac{F(\wi,\wo, \eta)G(\wi,\wo, \alpha)D(\wi,\wo, \alpha)}{4 (\wi \cdot n) (\wo \cdot n)},
\end{equation}
where the set of parameters $p = \left\{ \kd, \eta, \alpha \right\}$ is to be fit per material, with $\kd \in [0,1]^3$ being the diffuse albedo of the Lambertian term, $\eta \in [1,10]^3$ corresponds to a colored index of refraction and model how the light is reflected with the Fresnel term $F$, and $G$ and $D$ are respectively the geometric attenuation and normal distribution terms of the microfacet model that depend on a roughness parameter $\alpha \in [0,1]^2$ for anisotropic materials. 

We choose this analytical formulation for its simplicity and its widespread use in real-time rendering. The GGX model is a good fallback for our hybrid model because it is a good trade-off between quality and performance. As shown in Table \ref{tab:render_latency}, it is faster to evaluate than any MLP and can also represent a wide variety of materials. In practice, we observe that across the 312 BRDFs fitted  in our experiments, this single-lobe model combined with our neural correction is sufficient to reproduce the vast majority of target appearances accurately.

\paragraph*{Network Architecture.} As illustrated in Figure~\ref{fig:method}, we model both $\fc$ and $\fg$ with one shallow MLP $\nn_\p$. In practice, we use an MLP with a few hidden layers (1-3) and with limited hidden dimensions (16-32)
because such architectures can be evaluated in a fragment shader during the rendering stage for real-time applications \cite{weinreich2024real}, and optimized leveraging cooperative vectors \cite{laurHardw25}. The inputs of $\nn_\p$ are the incoming and outgoing directions $\wi,\wo$ stored in Cartesian coordinates and the latent code $z$ for the selected BRDF. The last linear layer is split into two parts to which we apply an exponential activation for the residual term $\fc$ and a sigmoid activation for the gating term $\fg$.

\begin{figure}[t]
\includegraphics[width=\linewidth]{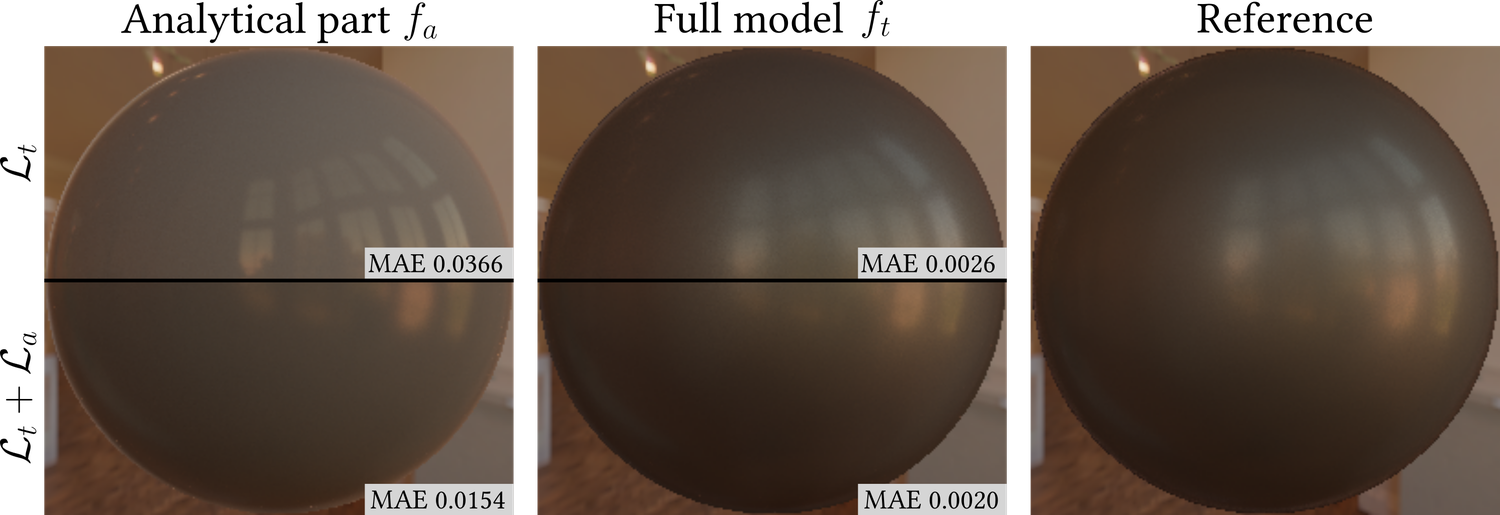}
\caption{Impact of the analytical loss $\mathcal{L}_a$ term. 
\textbf{Top-part}: with full loss~\eqref{eq:loss}, $\fa$ remains close to the target reflectance. 
\textbf{Bottom-part}: without $\mathcal{L}_a$, $\fa$ is allowed to deviate from the target. This impacts the overall quality and weakens the editability, as the analytic parameters $p$ no longer match the final appearance.
Tested Material: \texttt{alumn-bronze} from MERL. }
\label{fig:loss_analytic}
\end{figure}

\subsection{Fitting Method\label{sec:training}}
We now describe how the parameters of our hybrid BRDF model are fitted to measured BRDF data.  
The parameters describing our models consist of the latent code $z\in\R^l$, the analytical parameters $p$ and the neural network weights $\p\in\R^m$. Although it would be possible to fit all these parameters independently for each target BRDF, this would not be practical in a real-time rendering context, as it would require storing a dedicated set of network weights $\p$ for every material. To avoid this prohibitive cost, our fitting strategy relies on sharing the network weights $\p$ across a collection of BRDFs $\{\ft^i\}_{i=1}^N$. %

In practice, we rely on a mini-batch stochastic gradient descent to jointly optimize the common neural network weights $\p$ together with the per-BRDF analytical parameters $\mathbf{p} = \{p_i\}$ and the per-BRDF latent codes $\mathbf{z} = \{z_i\}$.
At each training iteration, we randomly sample a batch $\mathcal{B} = \{(\omega_i, \omega_o)\}$ of 1024 incident and outgoing direction pairs drawn from a cosine-weighted distribution over the hemisphere.  
Each BRDF $f^i$ in the dataset is evaluated at these directions, and we minimize the following objective:  
\begin{equation}
\label{eq:loss}
\sum_{i=1}^{N} \sum_{\omega \in \mathcal{B}}
\mathcal{L}_a\!\left(f^i(\omega), f_a(\omega; p_i)\right)
+
\mathcal{L}_{t}\!\left(f^i(\omega), f_t(\omega; p_i, z_i, \p)\right),
\end{equation}
where $f^i$, $f_a$, and $f_t$ denote the reference, analytical, and hybrid BRDFs, respectively. 
The analytical loss term $\mathcal{L}_a$ encourages the analytical component to approximate the target BRDF as accurately as possible on its own, whereas the hybrid loss term $\mathcal{L}_t$ supervises the full model.  
Including both terms in the objective is essential to obtain the desired behavior: the analytical model captures the primary perceptual characteristics of the material, such as the dominant diffuse and specular lobes, whereas the neural network focuses on modeling complex residual effects that cannot be well approximated by the analytical formulation alone. Figure~\ref{fig:loss_analytic} demonstrates this behavior on an example material with the result obtained with the combined loss \eqref{eq:loss} compared to the result obtained with solely the hybrid loss term $\mathcal{L}_{t}$. 
Both loss terms use the same error metric:
\begin{equation}
\mathcal{L}_{a,t}(f_1, f_2)
=
\left\|
\log\!\left(1 + \cos(\theta_i)\, f_1 \right)
-
\log\!\left(1 + \cos(\theta_i)\, f_2 \right)
\right\|^2,
\end{equation}
where $\theta_i$ denotes the incident elevation angle.  
The logarithmic compression reduces the dynamic range%
while the cosine elevation downweights high-energy grazing-angle samples, which are known to dominate the loss and destabilize training~\cite{lowBRDF12}.

In our implementation, the gradients are then computed using auto-differentiation from PyTorch~\cite{torch}, and optimization is done with the AdamW optimizer~\cite{adam, adamW}.
In practice, we optimize the parameters $\p$ of the neural network $\nn$ and the per BRDF parameters $z, p$ starting with a learning rate of $0.005$ and a cosine decay for $200$k steps. We find that clipping gradient norms over $0.01$ helps with stability. A full training for $100$ BRDFs (such as the MERL dataset) takes about 10 minutes on an RTX 5080.

\subsection{Rendering Implementation}
Integrating our hybrid model into a rendering system is straightforward.
At runtime, the analytical component is handled exactly as in a conventional renderer, with the only addition being the inclusion of the latent vector in the material definition.
The neural network is exported as a shader function that is evaluated after the analytical BRDF computation and outputs the gating and corrective terms, which are then used to update the analytical BRDF value according to Eq.~\eqref{eq:formulation}.
As a result, the memory overhead of the hybrid model is limited to the storage of the latent vector, while the size of the neural network (14~kB for a 32$\times$3 MLP shared between all BRDFs of the dataset) determines the performance overhead.
Both can be adjusted to meet the performance and quality requirements of the target rendering application.
Moreover, our hybrid model can be efficiently sampled by taking advantage of its analytical component~$\fa$.

\section{Results}
\label{sec:results}
\begin{figure*}[t]
  \centering\includegraphics[width=1.\textwidth]{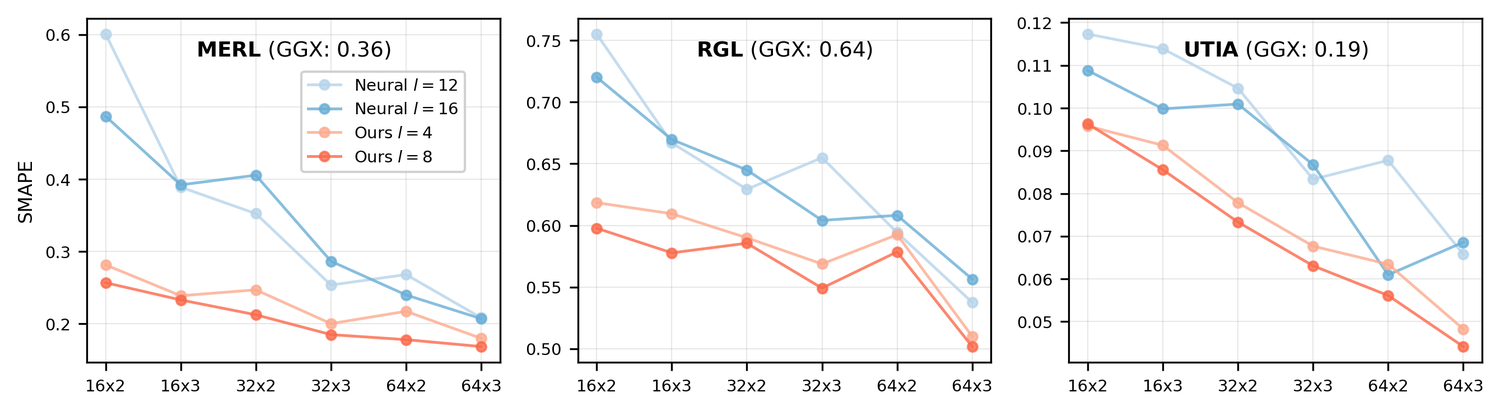}%
  \caption{Average SMAPE on BRDF reconstruction for the 3 datasets for multiple MLP size. 
  Since the analytical part of our hybrid model requires 7/8 parameters, we compare it to a fully neural model with 8 additional latent dimensions to provide a comparison with the same number of parameters (per BRDF) for both methods.
  The metric next to the dataset name is the SMAPE metric of the analytical part of the hybrid model on the particular dataset. 
  For the MERL and RGL datasets, our approach, for small-sized MLPs, outperforms the Neural model 
  or the GGX microfacet-based model. 
  The improvement is not as significant in the case of the UTIA dataset because its materials
  are very similar, and the anisotropic GGX model alone already provides a good approximation.}
  \label{fig:BRDF_result}
\end{figure*}

This section provides numerical results that demonstrate our model's ability to fit complex measured BRDFs (section~\ref{sec:quality}). Additionally, in section~\ref{sec:ablation}, we conduct ablation experiments that validate our hybrid architecture choice together with its capability of representing a wide range of BRDFs. Then we evaluate the raw rendering performance cost in Section~\ref{sec:renderingPerf}.

We evaluate our method on the following datasets: 100 Isotropic BRDFs from MERL~\shortcite{matuDatad03}, 51 Isotropic and 11 anisotropic BRDFs from the RGL~\shortcite{Dupuy2018Adaptive}, and 150 anisotropic BRDFs from the UTIA BRDF dataset \shortcite{filip14template}. These three datasets provide a diverse collection of measured real-world materials covering a broad range of appearances, such as fabrics, metals, plastics, paints, etc. 

In the experiments presented in this section, we trained our model on each dataset, using a single network shared by all BRDFs within that dataset. This highlights the ability of our hybrid model to represent a large variety of BRDF behaviors using a common network.

\begin{figure*}[ht]  
  \centering\includegraphics[width=0.9\textwidth]{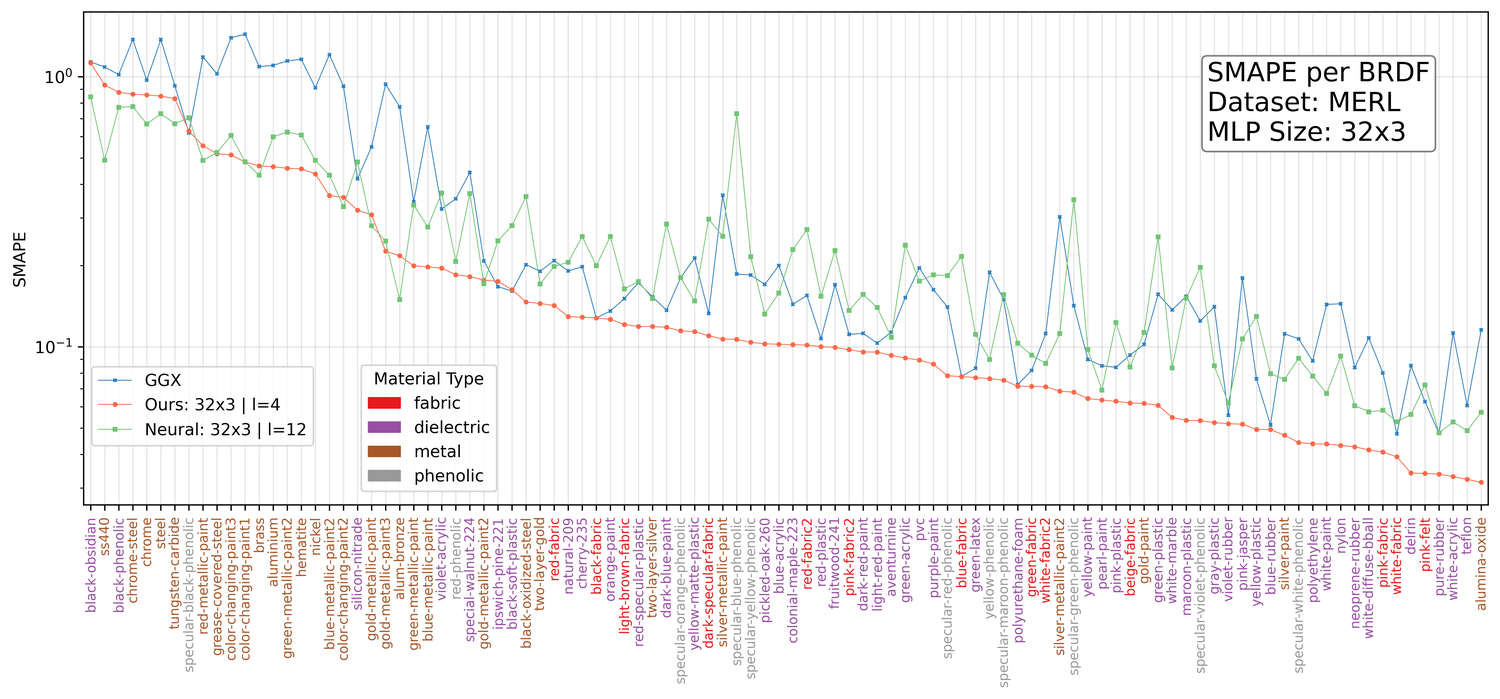}
  \caption{Per BRDF SMAPE on the MERL database, which consists of 100 measured materials. 
  Our model with the $32\times3$ MLP and a 4D latent space is compared to the Neural model~~\cite{zeltRealt24} with a $32\times3$ MLP and a 12D latent space, as well as a microfacet-based GGX model.
  Of the 100 materials, our hybrid model outperforms the Neural model for 88 materials 
  of them. The supplemental material provides the same analysis for the RGL and UTIA databases.}
  \label{fig:MERL_metrics}
\end{figure*}

\subsection{Quality Analysis}\label{sec:quality}
To assess the quality of our model, we use two complementary metrics: one measuring the BRDF reconstruction error in function space and the other measuring the rendering error against a reference scene. We then compare these values to the ones obtained with our implementation of the fully neural BRDF architecture from \cite{zeltRealt24}, and we refer to this method as \emph{neural} in the following. This architecture consists of a shallow MLP and a frame transform module that outputs 3 shading frames to transform the input directions. To be comparable, we replace the encoder part with a trainable latent code with a dimension that matches the dimension of the concatenation of our latent code $z$ and the parameters of the analytical model $p$. We therefore consider a latent space of dimension $12$ (resp. $16$) to compare with our approach, with a latent code of dimensions $4$ (resp. $8$) since we also require $7$ ($8$ for anisotropic) parameters for the analytical model.

\paragraph*{BRDF Reconstruction Quality.}
The BRDF-space reconstruction error is computed with the Symmetric Mean Absolute Percentage Error (SMAPE) 
by sampling both the target and the estimated BRDF with $10^{6}$ directions $\mathcal{B} = \left\{(\wi,\wo)\right\}$.
\begin{equation}
\SMAPE(f, \hat{f}) = \frac{2}{10^{6}}\!\sum_{\omega\in\mathcal{B}} \frac{|f(\omega) -\hat{f}(\omega) |}{|f(\omega)| + |\hat{f}(\omega) |}\,.
\end{equation}
Figure~\ref{fig:BRDF_result} and \ref{fig:MERL_metrics} show results using this metric. In particular, figure~\ref{fig:BRDF_result} presents the average SMAPE value obtained across all BRDFs of each tested dataset for various MLP sizes, both for our hybrid approach and the neural approach \cite{zeltRealt24}. This clearly shows that our hybrid approach provides better BRDF reconstruction, especially for smaller models. In practice, this means that our model can match the quality of fully neural models at much lower computational cost, resulting in a higher frame-rate in real-time applications or shorter rendering times in path-tracing. 
As the network size increases, the gap between the two approaches decreases. This is expected since larger networks imply better expressiveness and can approximate the BRDF sufficiently well on their own. The relative contribution of the analytical part is therefore less important. Note that this also explains the smaller difference observed on the UTIA dataset, where even very compact networks already achieve low SMAPE values (roughly $10\times$ smaller than on the other dataset).

\begin{figure}[ht]
  \includegraphics[width=\linewidth]{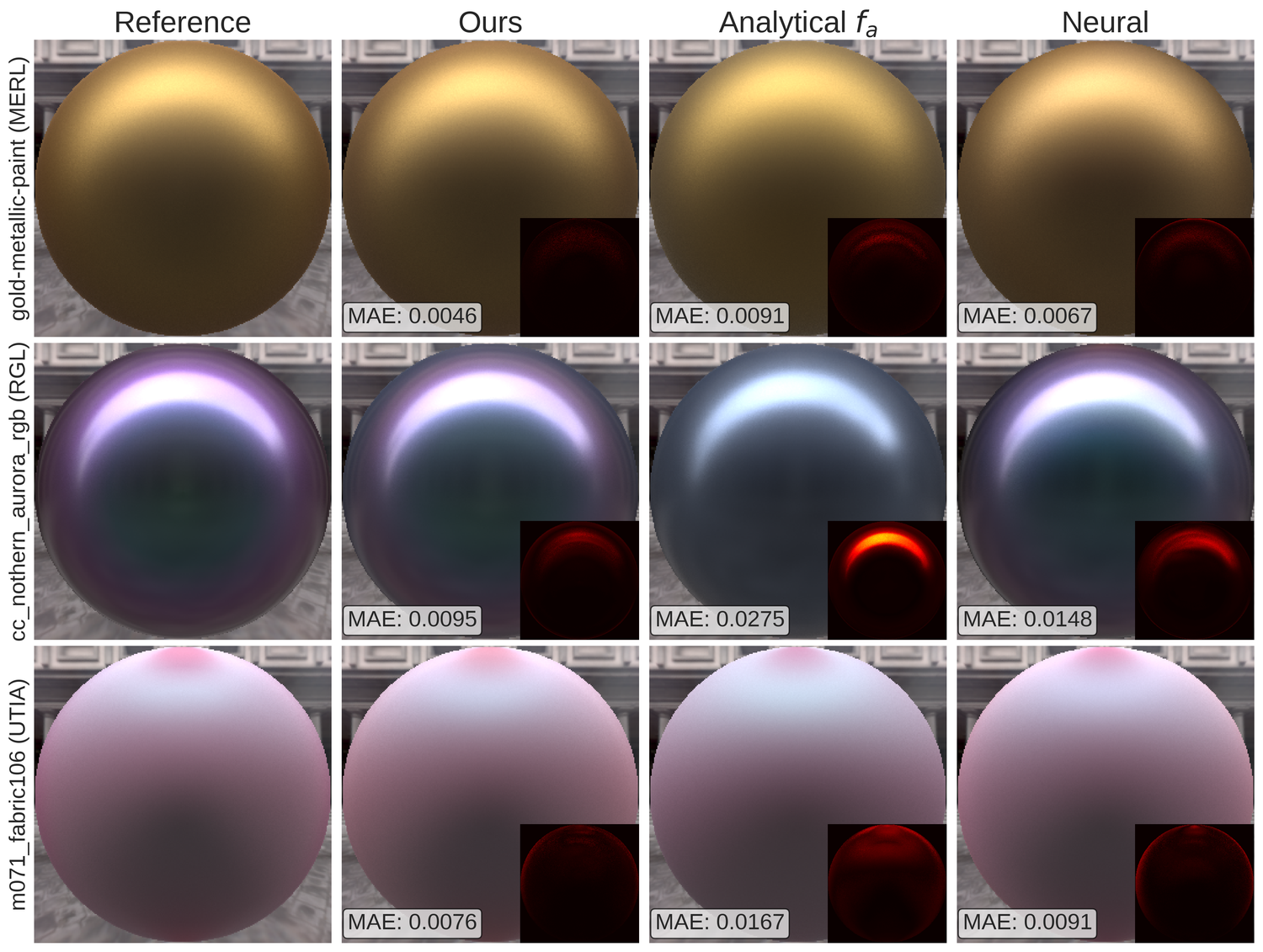}%
  \caption{Per-material rendering comparisons between our model, its underlying analytical component, and the neural model. The insets show the per-pixel absolute error as well as the color-averaged MAE for the whole image. 
  The number of parameters (12) and the MLP size ($32\,\times\,3$) are the same for the neural model and ours.
  }
  \label{fig:render_result}
\end{figure}

\paragraph*{Rendering Quality.} 
Based on the Mean Absolute Error (MAE), we evaluate our model by comparing its
renderings with those obtained with only its analytical component, and the neural model 
from~\cite{zeltRealt24}. These renderings are computed with 1024 samples per pixel (spp) and 
compared against a reference image computed with the measured BRDF.
Figure \ref{fig:render_result} shows results for one material of each of the considered datasets
in a simple test scene consisting of a sphere lit by the \texttt{Uffizi} environment map.
The number of parameters (12) and the MLP size (32x3) are the same for the neural model and ours.
As shown in the Figure~\ref{fig:render_result}, our  model produces a color-averaged MAE lower than the ones from the Neural or GGX models.
Table~\ref{tab:render_mae}, which gathers the results obtained for the 300 materials tested as well as different MLP sizes and parameters, confirms the performance of our model in the vast majority of cases, especially for small MLPs. In summary, with small MLPs ($16\times3$ and $32\times3$), our model consistently outperforms the neural model with the same network sizes, for a comparable rendering cost and identical memory usage. In practice, we find that the $32\times3$ network offers a good compromise between expressiveness and evaluation time, and therefore adopt this configuration for the major part of our rendering experiments.

\begin{table*}[ht]
\centering
\caption{Mean Absolute Error (MAE) in terms of rendering. For the three datasets, we compare
rendered images (same configuration as the ones from Fig.~\ref{fig:render_result}) between a GGX microfacet-based model, our hybrid model, and the Neural one for the same number of parameters and for different sizes of MLP. For a given dataset, the MAE is averaged over all renderings. Except for the RGL dataset and an MLP size of $64\times3$ or $16\times3$, our hybrid model exhibits better results.}
\label{tab:render_mae}
\resizebox{\textwidth}{!}{%
\setlength{\tabcolsep}{2.5pt}%
\begin{tabular}{|c|cc|cc|cc|cc|cc|cc|}
\hline
\multirow{3}{*}{MLP}& \multicolumn{4}{c|}{\textbf{MERL} (GGX MAE: 0.0099)} & \multicolumn{4}{c|}{\textbf{RGL} (GGX MAE: 0.0333)} & \multicolumn{4}{c|}{\textbf{UTIA} (GGX MAE: 0.0201)} \\\cline{2-13}
 & \multicolumn{2}{c|}{12 params} & \multicolumn{2}{c|}{16 params} & \multicolumn{2}{c|}{12 params} & \multicolumn{2}{c|}{16 params} & \multicolumn{2}{c|}{12 params} & \multicolumn{2}{c|}{16 params} \\\cline{2-13}%
 & Ours \latent{4} & Neural \latent{12} & Ours \latent{8} & Neural \latent{16}  & Ours \latent{4} & Neural \latent{12} & Ours \latent{8} & Neural \latent{16}  & Ours \latent{4}  & Neural \latent{12} & Ours \latent{8} & Neural \latent{16} \\
\hline
$16\times3$ & \cellcolor[RGB]{231,244,202} 0.0059 & \cellcolor[RGB]{219,153,168} 0.0086 & \cellcolor[RGB]{214,237,194} 0.0056  & \cellcolor[RGB]{227,161,168} 0.0084 & \cellcolor[RGB]{244,184,174} 0.0267 & \cellcolor[RGB]{251,201,182} 0.0256  & \cellcolor[RGB]{254,222,191} 0.0245 & \cellcolor[RGB]{219,153,168} 0.0288 & \cellcolor[RGB]{254,249,220} 0.0109 & \cellcolor[RGB]{219,153,168} 0.0123 & \cellcolor[RGB]{243,250,214} 0.0106 & \cellcolor[RGB]{243,182,173} 0.0119 \\
$32\times3$ & \cellcolor[RGB]{178,221,188} 0.0051 & \cellcolor[RGB]{241,249,211} 0.0062 & \cellcolor[RGB]{186,225,190} 0.0052 & \cellcolor[RGB]{254,250,221} 0.0067 & \cellcolor[RGB]{241,249,211} 0.0203 & \cellcolor[RGB]{254,232,200} 0.0238 & \cellcolor[RGB]{241,249,211} 0.0203 & \cellcolor[RGB]{246,251,217} 0.0208 & \cellcolor[RGB]{211,236,194} 0.0101 & \cellcolor[RGB]{244,250,215} 0.0106 & \cellcolor[RGB]{186,225,190} 0.0098 & \cellcolor[RGB]{254,227,196} 0.0113 \\
$64\times3$ & \cellcolor[RGB]{157,202,179} 0.0046 & \cellcolor[RGB]{168,216,186} 0.0049 & \cellcolor[RGB]{153,194,175} 0.0044 & \cellcolor[RGB]{161,210,183} 0.0048 & \cellcolor[RGB]{156,201,178} 0.0149 & \cellcolor[RGB]{153,194,175} 0.0143 & \cellcolor[RGB]{168,216,186} 0.0160 & \cellcolor[RGB]{159,207,181} 0.0153 & \cellcolor[RGB]{156,201,178} 0.0094  & \cellcolor[RGB]{185,224,190} 0.0098 & \cellcolor[RGB]{153,194,175} 0.0093 & \cellcolor[RGB]{202,232,193} 0.0100 \\
\hline
\end{tabular}%
}
\end{table*}

\subsection{Ablation Studies}
\label{sec:ablation}

\paragraph*{Model Architecture Variants.}
To validate our choice of the split additive and multiplicative terms for the model architecture~\eqref{eq:formulation}, we ran ablations to compare it to three variants:
\begin{itemize}
    \item \emph{Additive} term only: the MLP has a single output with an identity activation that is directly added to the analytical model.
    \item \emph{Log Multiplicative}: the MLP has a single output with an identity activation that is multiplied by the analytical model in logarithmic scale, and the output is then exponentiated. This approach ensures that the MLP's output range can match the range of the analytical BRDF values, making the comparison fairer.
    \item \emph{Analytical Input}: the same architecture as our model is used, but the MLP takes additionally the analytical model in logarithmic scale as input.
\end{itemize}

\begin{table}[ht]
\centering
\caption{BRDF-space SMAPE metric for architecture variations of our model. The values are computed for a $32\times3$ MLP.}
\label{tab:ablations}
\begin{tabular}{|c|c|c|c|}
\hline
 \textbf{Variant} & \textbf{MERL} & \textbf{RGL} & \textbf{UTIA} \\
\hline
Analytical Input & 0.2015 & 0.5493 & 0.0621 \\
Additive & 0.4231 & 0.7148 & 0.0842 \\
Log Multiplicative & 0.2186 & 0.5907 & 0.0708 \\
Ours & 0.1974 & 0.5446 & 0.0644 \\
\hline
\end{tabular}
\end{table}

The results in Table~\ref{tab:ablations} show that using both additive and multiplicative terms together yields better performance. Note that using only an additive model significantly degrades the results. Furthermore, we do not observe any notable difference when adding the analytical input. For simplicity, our model omits the analytical input from our model.

\paragraph*{Sparse Training.}
It is possible to add new materials after training on a certain number of BRDFs without needing to update the weights of the MLP $\nn_\p$. We train on half the materials from the MERL Dataset, stratified by material type (metallic, fabric, dielectric, phenolic), and then optimize $z$ and $p$ only for the remaining materials. We compare in Table~\ref{tab:reducedDataset} the performance between a full training and a sparse training, differentiating between materials that $\nn$ were trained on and the others.
When training on only 50\% of the MERL BRDFs, we observe a moderate degradation in BRDF-space metrics compared to training on the full dataset. Interestingly, the impact on 
image space metrics remains limited, with the MAE increasing only slightly from 0.0087 to 0.0099. When evaluating separately on the training subset, the performance remains nearly unchanged between full and sparse training, confirming that optimizing only the latent parameters $(z,p)$ is sufficient to represent materials accurately already seen during training. 
On the held-out subset, the degradation is more noticeable in BRDF-space, but remains moderate in render-space. This suggests that the network $\nn_\p$ captures a material prior that generalizes well to unseen BRDFs, even when trained on only half of the dataset. 
Overall, these results indicate that new materials can be incorporated after training without updating the weights of $\nn_\p$, while maintaining high rendering fidelity. Additional results and renders are provided in the supplemental material.

\subsection{Raw Rendering Performance Evaluation}
\label{sec:renderingPerf}

We measure the impact of evaluating an MLP on the BRDF evaluation in BRDFExplorer~\cite{DisneyBRDFExplorer}.
Table~\ref{tab:render_latency} reports the frame rendering time when using the analytical BRDF alone, our hybrid BRDF, and the neural BRDF baseline for various MLP sizes.
The evaluation time of our approach is comparable to that of the Neural model from \cite{zeltRealt24} at equal network size. 
However, as shown in the previous section, our model achieves significantly better reconstruction quality for the same MLP configurations. Moreover, as we show in the following section, our hybrid formulation enables efficient fallback mechanisms and importance sampling, making it particularly well-suited for real‑time rendering applications.

\begin{table}[t]
\setlength{\tabcolsep}{3.8pt}%
\centering
\caption{Results when using 50\% of the MERL BRDFs on training. For fairness, 50\% 
of BRDFs are not selected randomly from the entire dataset, but from within each material type (Figure~\ref{fig:MERL_metrics}). The values are computed for a $32\times3$ MLP.}
\label{tab:reducedDataset}
\begin{tabular}{|c|c|c|c|c|c|}
\hline
\multirow{2}{*}{\shortstack{\textbf{BRDF}\\\textbf{in training}}} & \multirow{2}{*}{\textbf{Subset}} & \multicolumn{2}{c|}{\textbf{BRDF-space}} & \multicolumn{2}{c|}{\textbf{Render-space}} \\\cline{3-6}
  & & MAE & SMAPE & MAE & PSNR \\
\hline
 100\% & \multirow{2}{*}{All} & 0.0031 & 0.1759 & 0.0087 & 37.594\\
  50\% && 0.0041 & 0.2104 & 0.0099 & 36.659\\\hline
  100\% & \multirow{2}{*}{\shortstack{Train\\subset}}& 0.0031 & 0.2057 & 0.0086 & 37.698\\
  50\% && 0.0032 & 0.2049 & 0.0088 & 37.530\\\hline
  100\% & \multirow{2}{*}{\shortstack{Held-out\\subset}} & 0.0031 & 0.1461 & 0.0893 & 37.491\\
  50\% && 0.0050 & 0.2160 & 0.0110 & 35.788 \\
\hline
\end{tabular}
\end{table}

\begin{table}[t]
\setlength{\tabcolsep}{2.5pt}
\centering
\caption{
Evaluation time (ms) of our hybrid model and the neural baseline 
for different MLP sizes, implemented in a GLSL shader in \emph{BRDFExplorer}
under environment map lighting, averaged over 100 spp, without importance sampling. Timings were recorded using NVIDIA Nsight on a RTX 5080. 
}
\label{tab:render_latency}
\begin{tabular}{|l|cccccc|c|}
\hline
\textbf{MLP Size}      & 16$\times$2 & 16$\times$3 & 32$\times$2 & 32$\times$3 & 64$\times$2 & 64$\times$3 & GGX \\ \hline
\textbf{Hybrid} (ms)   & 0.10        & 0.13        & 0.22        & 0.37        & 0.78        & 1.40       & \multirow{2}{*}{0.03} \\
\textbf{Neural} (ms)   & 0.16 & 0.17 & 0.30 & 0.45 & 0.98 & 1.59 &    \\ \hline
\end{tabular}
\end{table}

\section{Advantages of Our Hybrid Formulation\label{sec:discussion}}
\label{sec:discussion}

\begin{figure}[t]
  \includegraphics[width=\linewidth]{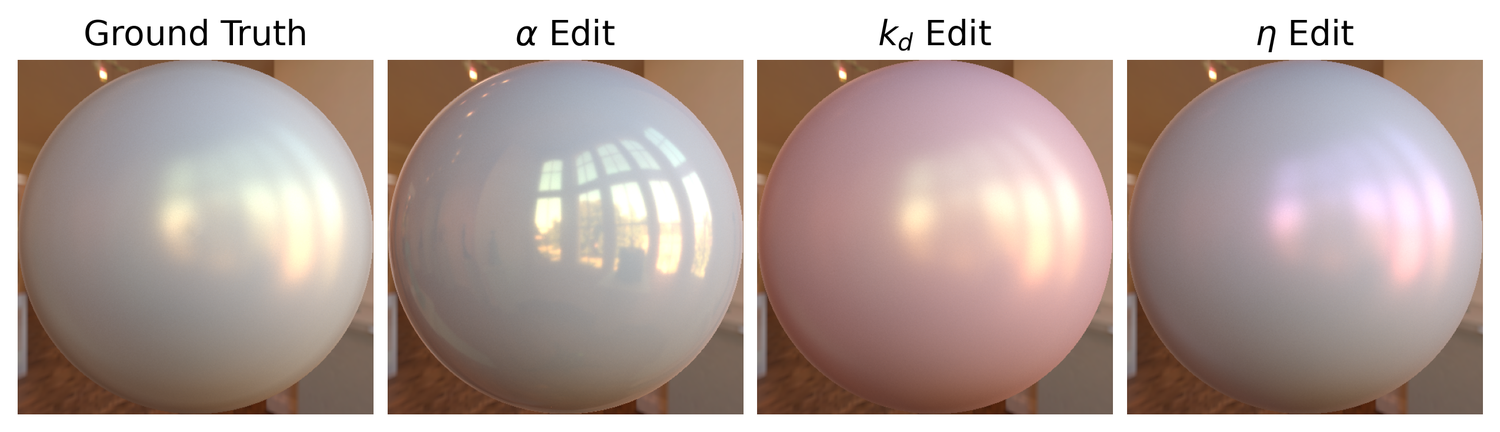}
  \vspace{-0.6cm}
  \caption{Editing the \texttt{satin-gold} BRDF from the RGL dataset by modifying the fitted parameters
  (roughness $\alpha$, albedo $k_d$, IOR $\eta$) of the analytical component of our model. 
  }
  \label{fig:all_edit}
\end{figure}

This section discusses the advantages of our proposed hybrid formulation. 
The analytical fallback naturally provides intuitive post-training authoring through analytical parameters (Section~\ref{sec:editability}), as well as an importance sampling strategy that improves convergence (Section~\ref{sec:importance_sampling}).
Also, our formulation allows for a selective evaluation of the neural network, which is beneficial in many real-time rendering scenarios (Section~\ref{sec:selective_eval}).

\subsection{Material Exploration}
\label{sec:editability}

\begin{figure*}[ht]
  \centering
  \includegraphics[width=0.9\linewidth]{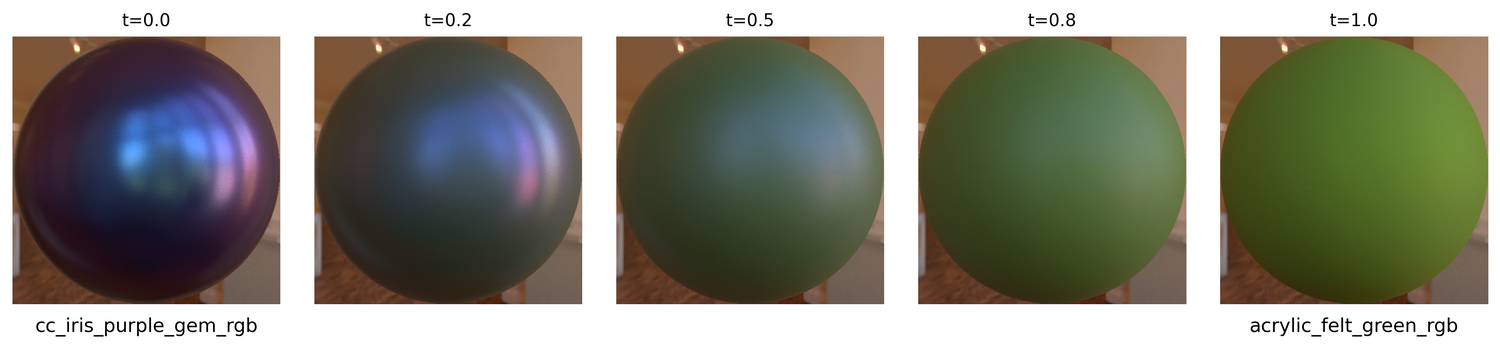}
  \vspace*{-0.25cm} 
  \caption{BRDF interpolation example. \textbf{From left to right}, we linearly interpolate (parameter $t$) between two materials from %
  the RGL dataset, the latent vector, and the analytical parameters of our hybrid model, showing very plausible results.}
  \label{fig:interpolation_MERL}
\end{figure*}

Since the neural component acts as a correction over the analytical model, modifying the analytical parameters produces consistent new appearances. Figure~\ref{fig:all_edit} shows examples of varying diffuse color, refractive index, and roughness on a fitted material. 
The results remain coherent throughout, which we attribute to the joint training strategy: the analytical loss ensures the GGX component encodes the dominant reflectance behavior, so the network only captures residual effects. 
As a consequence, the analytical parameters retain their physical meaning after fitting, and the network's correction adapts to moderate parameter changes.
This same reasoning explains why linear interpolation between two fitted materials produces plausible intermediate appearances (Figure~\ref{fig:interpolation_MERL}): blending both the analytical parameters and latent codes amounts to interpolating between two physically grounded baselines, with the network filling in consistent residuals throughout. 
We note that this is not guaranteed for large excursions in parameter space, as disentanglement between the analytical parameters and the latent code is not enforced. 
Nevertheless, within a reasonable range, the hybrid formulation provides a natural and stable basis for material exploration and spatially varying BRDF maps, where per-texel parameters and latent code maps can drive smooth appearance variation across a surface after training.

\subsection{Importance Sampling}
\label{sec:importance_sampling}

\begin{figure*}[ht]
\centering\includegraphics[width=0.9\linewidth]{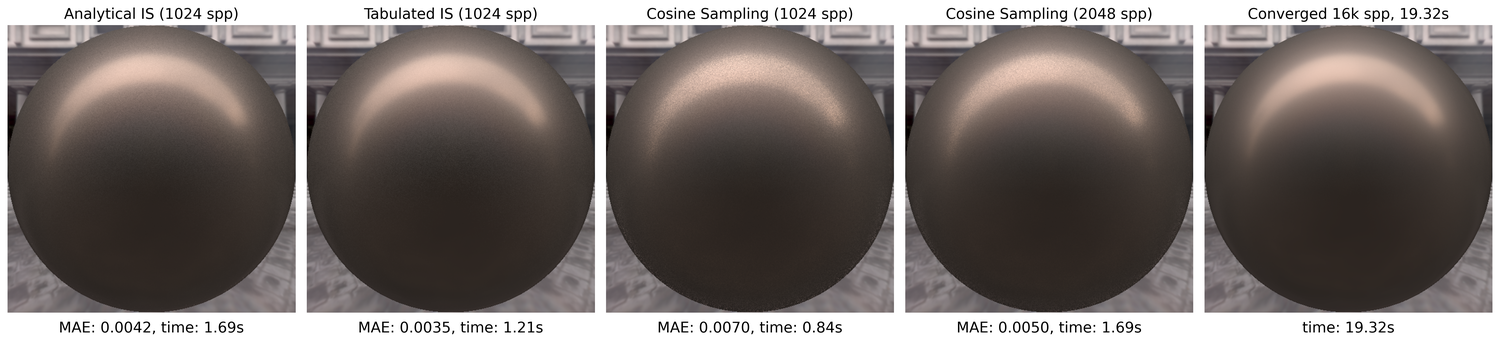}\\
  \caption{Importance Sampling (IS) comparisons in image space for the \texttt{red-phenolic} material (MERL). \textbf{From left to right:} IS based on the GGX distribution and a cosine lobe distribution, tabulated IS (similarly to~\cite{Lawrence04}), with a memory footprint of 5.62 MB, the cosine lobe IS, and a converged result using tabulated IS. 
  For all images, the BRDF values are taken from our fitted hybrid model.
  As shown by the two images on the left, using an importance sampling based on the microfacet model is not as optimal as the tabulated form but provides clear visual improvements over cosine lobe importance sampling. Additional comparisons are provided in the supplemental material.}
  \label{fig:IS_comp_alum_bronze}
\end{figure*}

\begin{figure}[ht]
\centering
  \includegraphics[width=1\linewidth]{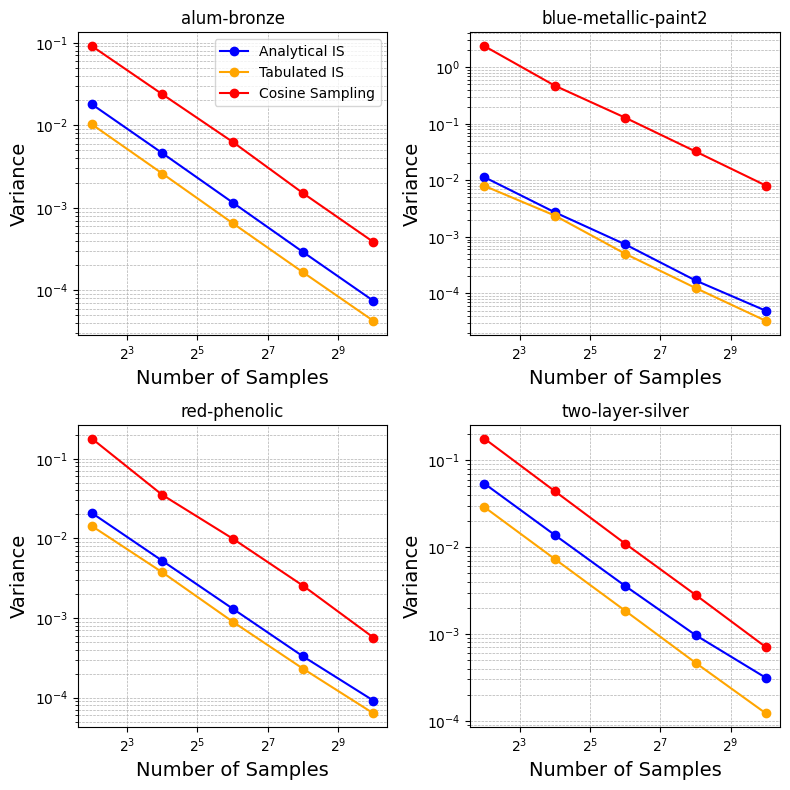}%
  \caption{Color-averaged variance of the absolute difference between an image rendered with importance sampling and a reference image computed with 16k samples. Analytical Importance Sampling  corresponds to applying  either cosine sampling or GGX sampling using a MIS procedure 
  for our model. As shown above, using our IS method approximates the true PDF quite accurately
  compared to the reference solution provided by tabulated Importance Sampling. 
  \vspace*{-10pt}
  }
  \label{fig:IS_variance}
\end{figure}

Efficient importance sampling remains an open challenge for neural BRDFs~\cite{xuComp25}. 
Existing solutions rely on evaluating a dedicated network at every ray bounce, either to directly predict sampling distributions~\cite{Bai2023} or to construct an analytical proxy~\cite{zeltRealt24}, introducing additional inference overhead and preprocessing cost. 
Our hybrid formulation provides a natural and overhead-free alternative: since the analytical component is jointly trained to remain close to the target BRDF, it serves directly as a sampling proxy. 
In practice, we apply multiple importance sampling (MIS) between cosine sampling and GGX lobe sampling without requiring any additional neural evaluation.
We compare our approach against a tabulated reference method~\cite{Lawrence04} sampled at 1-degree angular resolution (5.62 MB per material), which provides an upper bound on sampling efficiency. 
As shown in Figures~\ref{fig:IS_comp_alum_bronze} and~\ref{fig:IS_variance}, our method consistently outperforms cosine sampling and approaches the tabulated reference, at a fraction of the memory cost. 
This shows that the analytical component carries most of the BRDF energy, and it is therefore sound to use it as the primary source for importance sampling.

\subsection{Selective Neural Evaluation}
\label{sec:selective_eval}
\begin{figure*}[ht]
  \includegraphics[width=\linewidth]{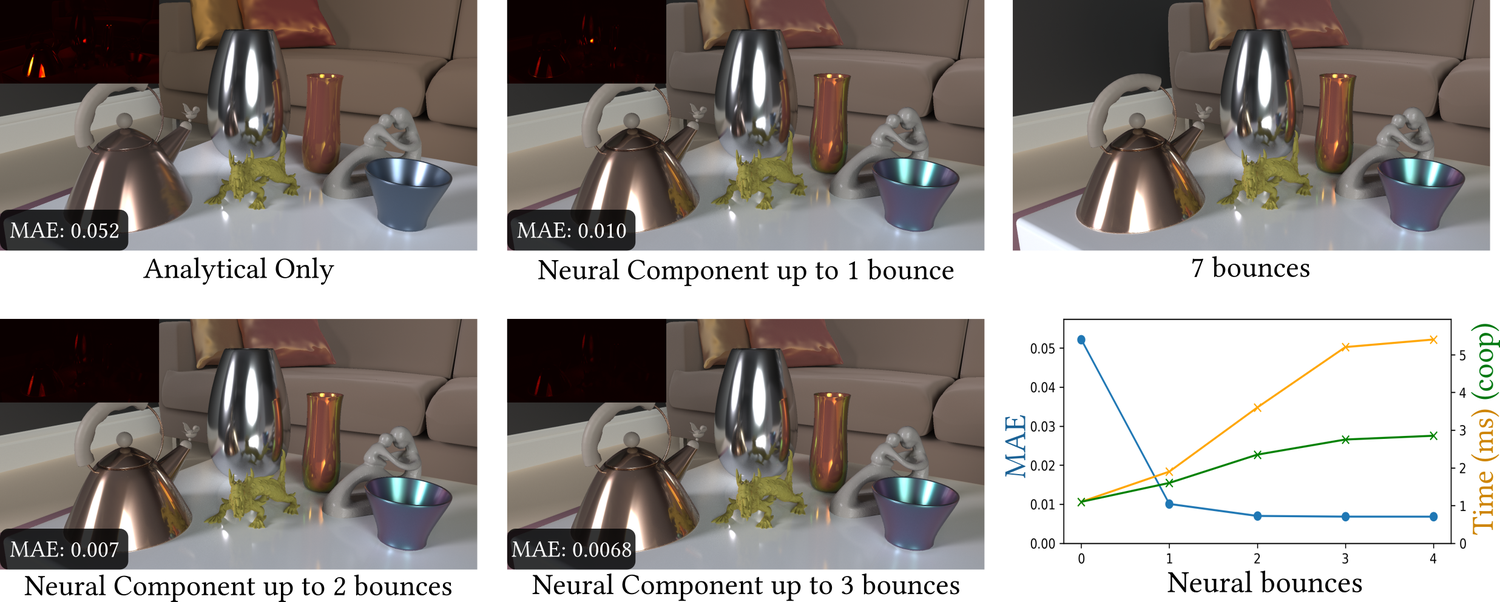}
  \caption{Path-Tracing performance evaluation. All materials of the scene's objects are rendered using our hybrid microfacet-neural model, which is fitted from three datasets. We show the effect in terms of MAE and rendering time (green and yellow curves show the time per sample with and without cooperative vector acceleration respectively), of falling back on the analytical component model after a certain number of bounces. The different materials used for this scene are listed in the supplementary material.}
  \label{fig:pt}
\end{figure*}

Our hybrid model improves visual fidelity over the analytical GGX model but introduces an inference overhead at every shading point.
Several strategies can mitigate this cost: leveraging hardware accelerators via the cooperative vectors extension~\cite{zeltRealt24}, exploiting low-bit integer arithmetic~\cite{xuComp25}, or amortizing inference temporally across frames~\cite{xuMobile26}. 
These are all orthogonal to our method and directly applicable to our compact shared MLP.
Beyond these, our hybrid formulation enables an additional and uniquely effective strategy: evaluating the network only when it meaningfully contributes to the final appearance, and falling back to the analytical component otherwise. 
To demonstrate this, we render a 1080p frame using a GPU path tracer with seven bounces, progressively disabling the neural component of our hybrid model after increasing bounce depths. 
Table~\ref{tab:perf_bounces} reports rendering times with and without cooperative vectors and Figure~\ref{fig:pt} shows the corresponding visual results.
The neural correction has negligible visual impact beyond the first two bounces, and evaluating the network only at the first hit is often sufficient. 
Yet the computational gains are substantial: with cooperative vectors, rendering time drops from 2.94~ms to 2.35~ms and 1.6~ms; without hardware acceleration, from 5.57~ms to 3.6~ms and 1.9~ms, approaching the 1.1~ms analytical baseline.
Crucially, selective inference and hardware acceleration are complementary.
The gains from reducing the amount of neural evaluations are consistent regardless of the inference back end.
This makes selective inference an effective and hardware-agnostic strategy, particularly valuable on platforms lacking dedicated neural inference support. 
More advanced heuristics based on surface roughness, screen-space contribution, or level of detail could also be envisioned to further reduce the rendering overhead.

\begin{table}[ht]
\setlength{\tabcolsep}{3.8pt}%
\centering
\caption{Rendering time (ms) for different numbers of path tracing bounces in Figure~\ref{fig:pt}.
Neural inference, for our hybrid model, is progressively enabled per bounce. Results are shown
with inlined and hardware-accelerated (CoopVec) inference. In this rendering scenario, CoopVec are 1.74 times faster on average.}
\label{tab:perf_bounces}

\begin{tabular}{|c|c|c|c|c|}
\hline
\multirow{2}{*}{\shortstack{\textbf{GGX}\\\textbf{until}}} &
\multirow{2}{*}{\shortstack{\textbf{Neural inference}\\\textbf{enabled until}}} & \multirow{2}{*}{\textbf{inlined}} & \multirow{2}{*}{\textbf{CoopVec}} & \multirow{2}{*}{\textbf{Speed up}} \\
& & & & \\
\hline
\multirow{8}{*}{\shortstack{7\\bounces}}& 0 bounces  & 1.10 ms & ---  & ---\\
& 1 bounces  & 1.90 ms & 1.60 ms & $\times$1.19 \\
& 2 bounces  & 3.60 ms & 2.35 ms & $\times$1.53 \\
& 3 bounces  & 5.20 ms & 2.75 ms & $\times$1.89 \\
& 4 bounces & 5.40 ms & 2.85 ms & $\times$1.89 \\
& 5 bounces   & 5.50 ms & 2.90 ms & $\times$1.90 \\
& 6 bounces  & 5.55 ms & 2.93 ms & $\times$1.89 \\
&7 bounces  & 5.57 ms & 2.94 ms & $\times$1.89 \\
\hline
\end{tabular}
\end{table}

\section{Limitations and Future Work \label{sec:limitations}}

\paragraph*{Model Limitations.}
The analytical part of the model is intrinsically interpretable from the microfacet theory and provides artistic control and editability (Section~\ref{sec:editability}). However, the latent space that drives the corrective layer is not itself interpretable, and editing the latent vector after the optimization process is not guaranteed to preserve BRDF-like behavior.
Moreover, our optimization process does not enforce any physical constraints such as energy conservation or Helmholtz reciprocity. However, we empirically observe that our learned BRDFs remain close to reciprocal in most cases (i.e., absolute error $< 1e-4$), and that their integrated energy tends to be similar or slightly lower than the reference.
These limitations suggest interesting future directions: for instance, by adding physically based priors into the model or introducing regularization terms to better structure the latent space.

\paragraph*{SVBRDFs and Appearance Acquisition.}
Neural methods are known to perform well when fitting noisy data due to the stochastic nature of the training process, while analytical models perform well on very sparse data due to their built-in inductive bias. 
Our Hybrid model would therefore be particularly well suited for SVBRDF acquisition using real data from multi-view and multi-light setups, as such data is often noisy and sparse.
In practice the spatial variations would be encoded in texture maps of $z$ and $p$ in a similar manner to Zeltner et al.~\shortcite{zeltRealt24}. The maps could include aggressive quantization and clusterization of the analytical parameters, leveraging latent space to apply the best possible correction.

\begin{figure}[bt]
  \includegraphics[width=1.0\linewidth]{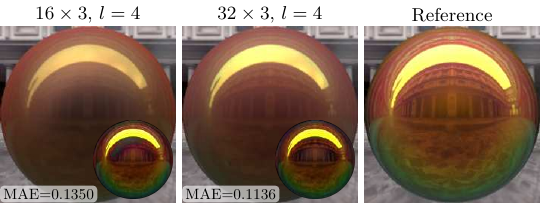}
  \vspace*{-0.7cm} 
  \caption{Our model struggles to  reproduce complex phenomena such as iridescence (from \texttt{cc\_amber\_citrine} from RGL) accurately without using a large neural network. At the bottom of each image is displayed, (\textbf{Left}) the color-averaged MAE and (\textbf{Right}) a map of the absolute error.}
  \label{fig:iridescent}
\end{figure}

\paragraph*{Iridescent Materials and Multi Lobe BRDFs.}
The current implementation of our model may struggle to accurately represent complex effects such as iridescence (Figure~\ref{fig:iridescent}). More generally, it tends to underperform on BRDFs that fall far outside the single-lobe framework with smooth angular variation. This is due to the strong single-lobe bias imposed by the GGX model used in  the analytical component (Section~\ref{sec:archi}). Capturing more complex effects would require increasing the expressive power of the model, either by enriching the analytical model or by relying on a more expressive neural component. Richer analytic models, such as multi‑lobe formulations, are significantly harder to optimize and introduce additional hyperparameters, while a larger neural component would allow to recover more complex appearances, at a substantial performance cost. In other words, finding the right balance between analytical priors and neural expressiveness remains an open challenge.

\section{Conclusion}
We introduced a hybrid reflectance model that combines an analytical microfacet-based GGX model with a lightweight neural network. 
The latter precisely and automatically fine-tunes the analytical model to reproduce appearances that are not representable by the analytical model alone.
We have shown that a shallow neural network, coupled with an analytical model, can reproduce complex appearances from various measured databases, while incurring only a slight increase in computational cost. 
Importantly, the analytical component remains a close approximation of the target appearance, keeping it meaningful to author or to use as a proxy.
Beyond reconstruction quality, our proposed hybrid formulation offers numerous practical advantages for real-world rendering workflows, including intuitive material authoring, efficient importance sampling without additional overhead and flexible performance optimizations through selective neural evaluation.
Overall, this results in a new practical reflectance model that jointly offers high expressiveness, stable behavior, fast evaluation, low memory footprint, and some editability, making it well-suited for production~use.

\section*{Acknowledgments}
This project was supported by the region Nouvelle-Aquitaine through the project VESPAA and by the project JRP 23IND14 xDDiff, that received funding from the European Partnership on Metrology, co-financed from the European Union’s Horizon Europe Research and Innovation Programme and by the Participating States.

\bibliographystyle{eg-alpha-doi} 
\bibliography{bibliography}

\end{document}